\documentclass[11pt]{article}
\usepackage{Moriond,epsfig}

\def\Journal#1#2#3#4{{#1} {\bf #2}, #3 (#4)}

\def\PLB{{\em Phys. Lett.}  B}

\def\ZPC{{\em Z. Phys.} C}
\def\EPJC{{\em Eur. Phys. J.} C}

\def\be{\begin{equation}}
\def\ee{\end{equation}}
\def\bea{\begin{eqnarray}}
\def\eea{\end{eqnarray}}

\begin{document}
\font\sfHUGE= cmss24  at 28truept
\font\sfhuge= cmss24  at 20truept
\font\sfXLARGE= cmss14  at 20truept
\font\sfLARGE= cmss14  at 14truept
\font\sfbig= cmss10  at 12truept
\font\sfmed= cmss10  at 10truept
\font\sfsml= cmss8   at  8truept
\setlength{\oddsidemargin}{0mm}
\setlength{\evensidemargin}{0mm}
\setlength{\topmargin}{0mm}
\setlength{\headheight}{0mm}
\setlength{\headsep}{0mm}
\setlength{\footskip}{0mm}
\setlength{\textheight}{202mm}
\setlength{\textwidth}{160mm}
\addtolength{\textheight}{68mm}
%
\addtolength{\textheight}{-68mm}
%
\begin{titlepage}
\thispagestyle{empty}
\begin{center}
{\sfHUGE MAX-PLANCK-INSTITUT F\"UR PHYSIK}\\[5mm]
{\sfhuge WERNER-HEISENBERG-INSTITUT}
\end{center}
\vspace*{2cm}
\begin{flushleft}
\Large MPI-PhE/2000-10\\
       May 15, 2000
\end{flushleft}
\vspace*{2cm}
\begin{center}
{\huge\bf
$\mathbf{\alpha_S}$ Evolution from 35 GeV to 202 GeV \\[2mm]
and Flavour Independence
}
\end{center}
\vspace*{20mm}
\begin{center}
\LARGE
O.~Biebel \\
\bigskip 
\bigskip
\Large
Max-Planck-Institut f\"ur Physik\\
80805 Munich, Germany
\end{center}

\vspace*{0pt\vfill}
\vfill
\begin{center}
{\sfXLARGE 80805 M\"unchen \hspace*{2mm}$\bullet$\hspace*{2mm} F\"ohringer Ring 6}
\end{center}
\vspace*{-15mm}
\end{titlepage}
\setlength{\oddsidemargin}{4mm}
\setlength{\evensidemargin}{4mm}
\setlength{\topmargin}{0mm}
\setlength{\headheight}{0mm}
\setlength{\headsep}{0mm}
\setlength{\footskip}{5mm}
\setlength{\textheight}{240mm}
\setlength{\textwidth}{160mm}
\setlength{\marginparwidth}{0mm}
\setlength{\marginparsep}{0mm}
\clearpage
%
%
%
\vspace*{-1cm}
\vbox to 21mm {
\hbox to \textwidth{ \hsize=\textwidth
\hspace*{0pt\hfill} 
\vbox{ \hsize=58mm
{
\hbox{ MPI-PhE/2000-10 \hss}
\hbox{ May 15, 2000\hss } 
}
}
}
}
\title{$\alpha_S$ EVOLUTION FROM 35 GEV TO 202 GEV AND FLAVOUR INDEPENDENCE}

\author{ O. BIEBEL }

\address{Max-Planck-Institut f\"ur Physik, F\"ohringer Ring 6,\\
80805 M\"unchen, Germany}

\maketitle\abstracts{
Determinations of the strong coupling constant $\alpha_S$ at centre-of-mass
energies of 192 through 202 GeV at LEP are presented. The energy evolution
of $\alpha_S$ is in agreement with the prediction of QCD. The combined
investigation of OPAL and JADE data in the energy range of 35 through 189 GeV
yields $\alpha_S(m_{\mathrm{Z}})=0.1187^{+0.0034}_{-0.0019}$. The strenght 
of the strong coupling is flavour independent if quark mass effects are taken 
into account. 
}

\section{Motivation}
Three fundamental properties follow from QCD: (i) the scale dependence of
the renormalised coupling strength, (ii) the flavour independence of the 
coupling apart from effects due to finite quark masses, and (iii) the scale 
dependence of the renormalised quark masses. It constitutes a significant
experimental test of QCD if the strong interaction obeys these properties.

The large range of energy covered by e$^+$e$^-$ colliders makes an investigation
of the energy evolution of $\alpha_S$ possible. Although the expected running of 
the coupling is more pronounced towards lower centre-of-mass energies, the uniform
analyses at LEP provide significant QCD tests up to the highest energies accessible.

\section{$\alpha_S(200 {\mathrm{~GeV}})$}
The excellent performance of the LEP collider in 1999 provided data at
$\sqrt{s}=192$, 196, 200, and 202~GeV comprising together about 220~pb$^{-1}$ 
per experiment. The results at each of the four $\sqrt{s}$ will be combined 
to derive the coupling strength at $\sqrt{s}=198$~GeV which is the luminosity 
weighted average energy.

Even though the selection of hadronic final states is trivial, background
contributions from essentially W pair production and large initial state
radiation have to be rejected. W pair events are excluded using the wide 
separation of the four jets and requiring di-jet masses different from the
W mass. The demand that the effective centre-of-mass energy of the hadronic 
final state is within 10-20\% of the nominal value is applied against initial 
state radiation events. These criteria select more than 75\% of the real 
hadronic final states and reduce the contribution from other processes
to less than 20\%.

From the selected events the observables thrust ($T$), heavy jet mass ($M_H$), 
total and wide jet broadening ($B_T$, $B_W$), $C$ parameter ($C$), and the 
value of the jet resolution parameter at the transition from 3 to 2 jets using 
the Durham jet algorithm ($y_3$) are measured. The strong coupling constant is 
found from fitting the QCD prediction, convolved with the hadronisation correction, 
to the distribution of an observable. The QCD prediction is the matched resummed 
next-to-leading logarithmic approximation (NLLA) with the full second order matrix 
element (${\cal O}(\alpha_S^2)$). The hadronisation correction is taken from various 
Monte Carlo event generators which although they were tuned to describe the data 
measured at 91~GeV yield a good representation of the data at energies of around 
200~GeV.

\begin{figure}
\centerline{\psfig{figure=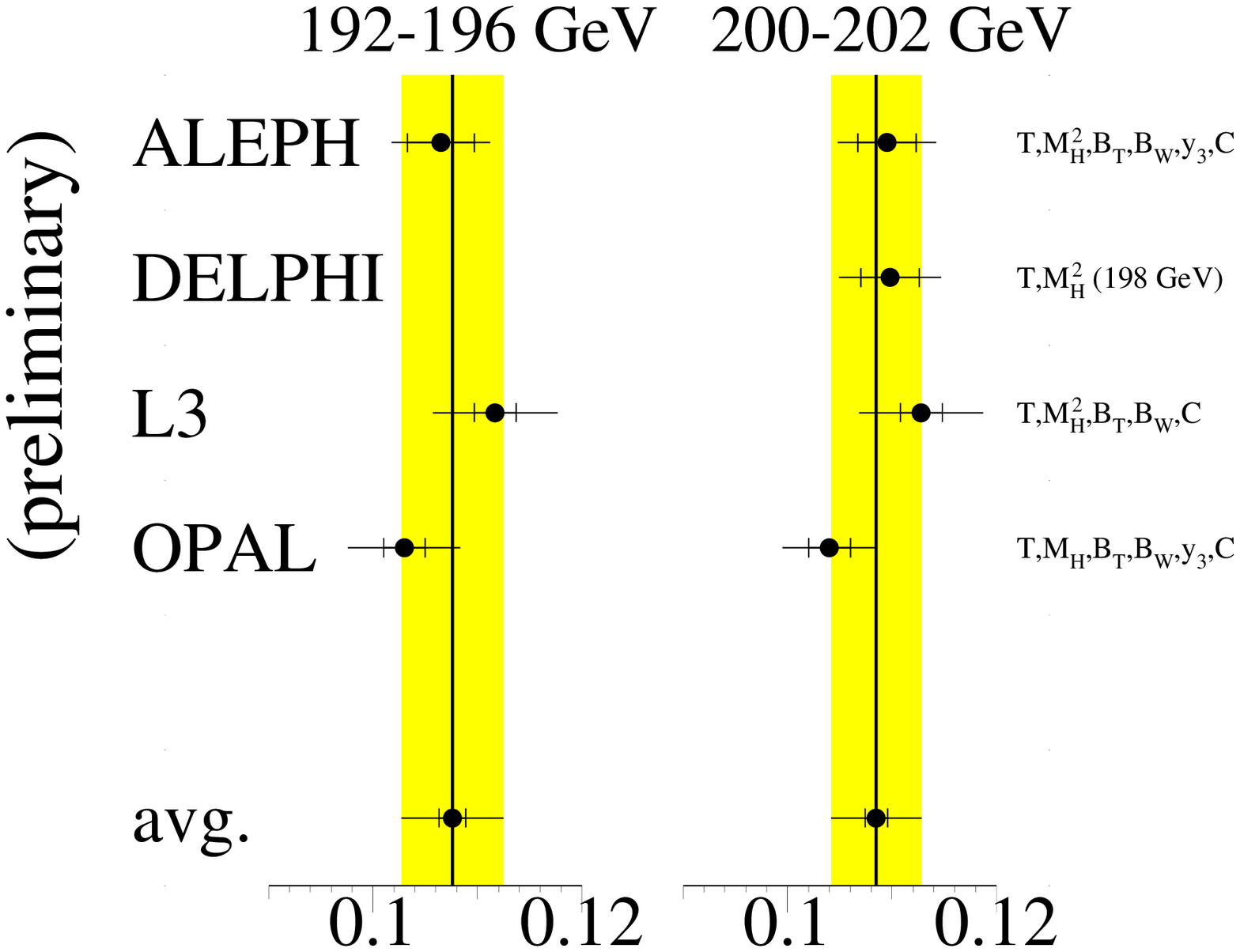,width=7cm} \hfill\psfig{figure=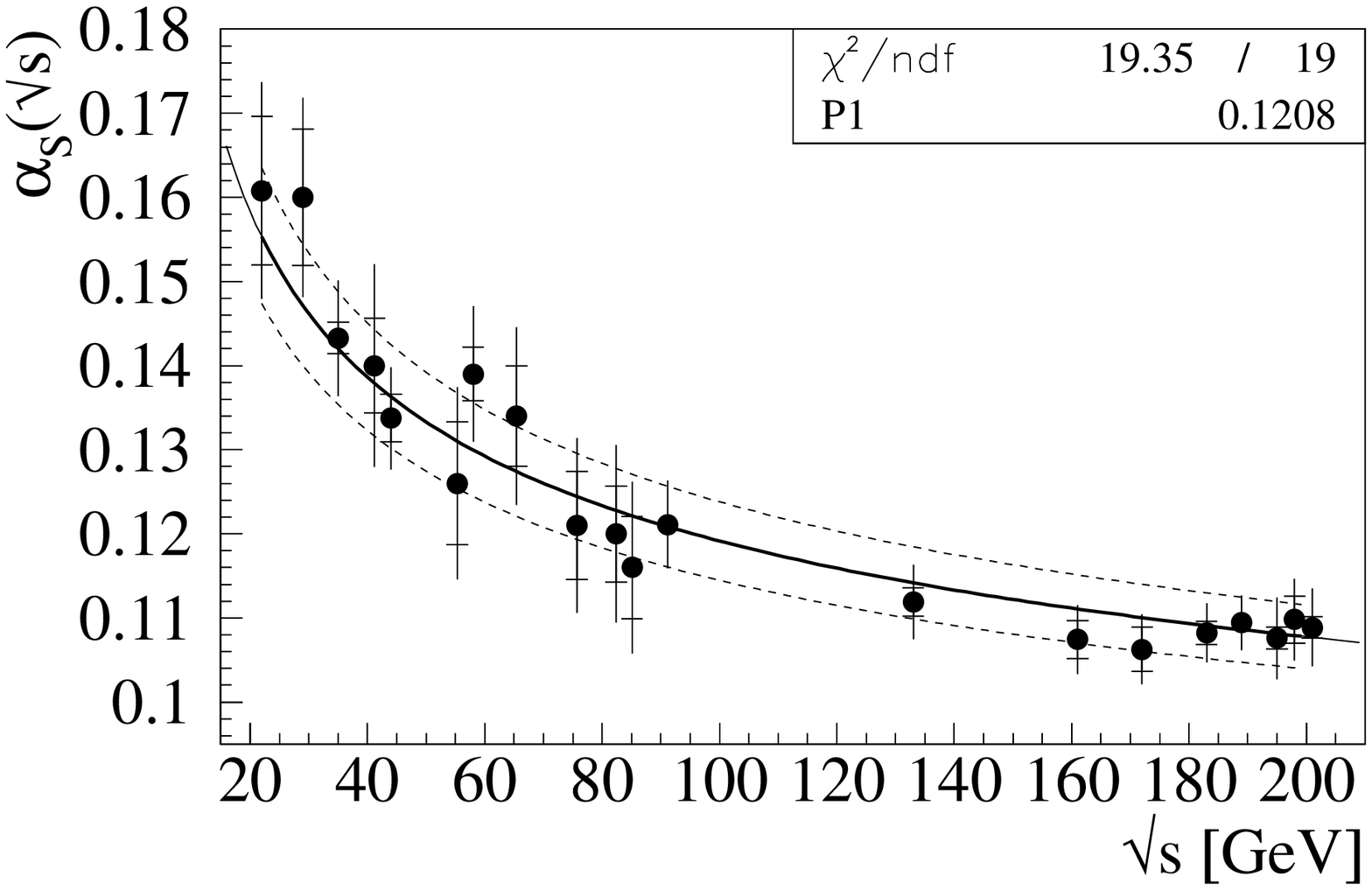,width=85mm}}
\caption{\label{fig-alphas}\label{fig-aslep2}
         Left: $\alpha_S$ results at the highest LEP energies. 
               The experimental contribution to the total errors is indicated. 
         Right: $\alpha_S$ between $\sqrt{s}=22$ and $202$ GeV, fitted with the
                4-loop prediction of the scale dependence.
}
\end{figure}

The left part of Fig.~\ref{fig-aslep2} shows the fit results of the LEP 
experiments~\cite{bib-aslep2} for $\alpha_S$ at $\sqrt{s}=192$-196 and 
200-202~GeV. Combining the values yields 
$\alpha_S(198 {\mathrm{~GeV}})=0.109 \pm 0.001_{\mathrm{(expt)}}
                                     \pm 0.005_{\mathrm{(theo)}}$.

\section{Energy evolution}
The LEP experiments contributed a large number of $\alpha_S$ determinations for 
$\sqrt{s} \ge m_{\mathrm{Z}}$ (see~\cite{bib-aslep2} and references therein). 
Exploiting initial and hard final state photon radiation, $\sqrt{s}$ as low as
30~GeV are accessible and have been investigated by L3 and DELPHI~\cite{bib-ISRFSR}.
These determinations of $\alpha_S$ below the Z mass are complemented by results of 
experiments at lower centre-of-mass energies~\cite{bib-TPCTOPAZ,bib-JADE} which,
even though already completed since long, re-analysed their data to employ the 
matched resummed NLLA and ${\cal O}(\alpha_S^2)$ predictions for an $\alpha_S$
determination. 

The right part of Fig.~\ref{fig-alphas} shows the $\alpha_S$ values versus the 
centre-of-mass energy.  The values with their experimental errors are fitted with 
the 4-loop formula for the scale dependence~\cite{bib-Ritbergen} of $\alpha_S$. To 
estimate the theory uncertainty the scale uncertainties of the single measurements 
are considered to be fully correlated. The fit yields $\alpha_S(m_{\mathrm{Z}}) = 
0.1208 \pm 0.0006_{\mathrm{(expt)}} \pm 0.0048_{\mathrm{(theo)}}$ which is not 
dominated by the single measurement at 91.2~GeV and which agrees with the
world average of $0.1184 \pm 0.0031$~\cite{bib-asworldavg}.

\begin{figure}
\centerline{\psfig{figure=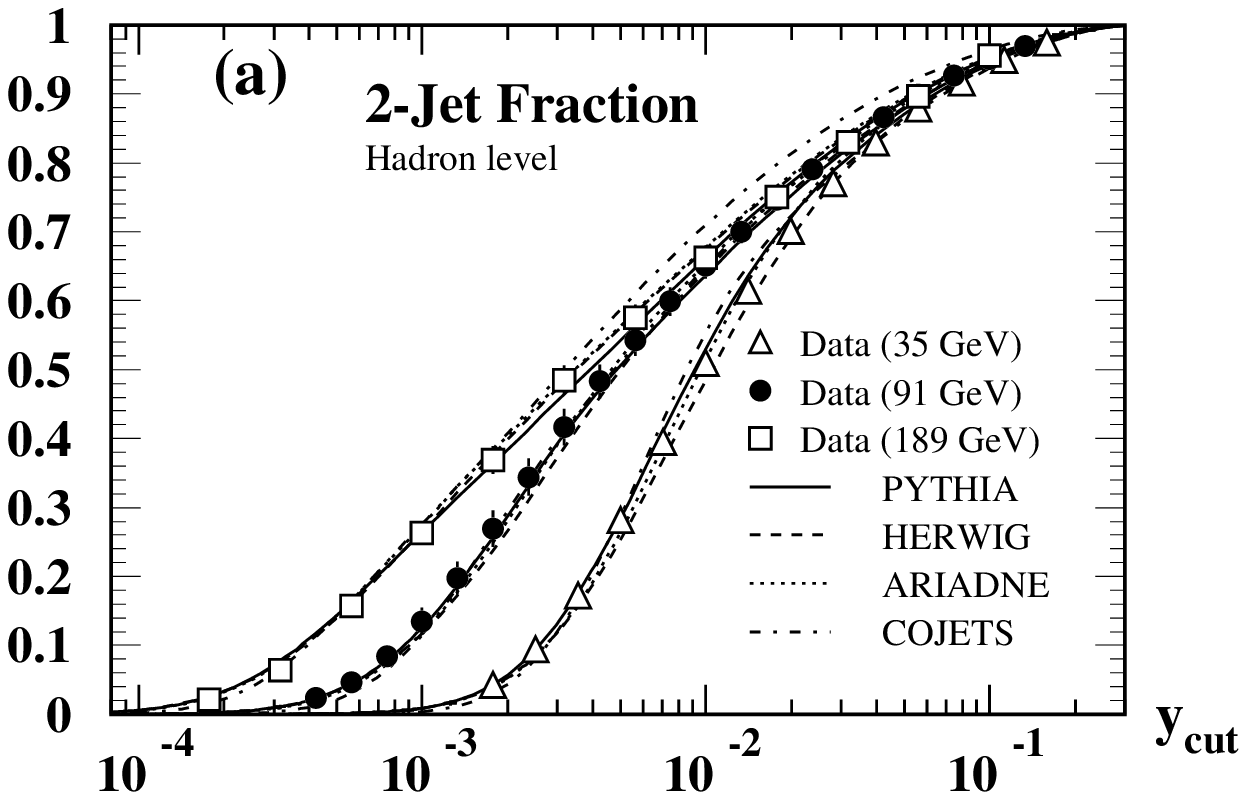,width=7cm} 
            \psfig{figure=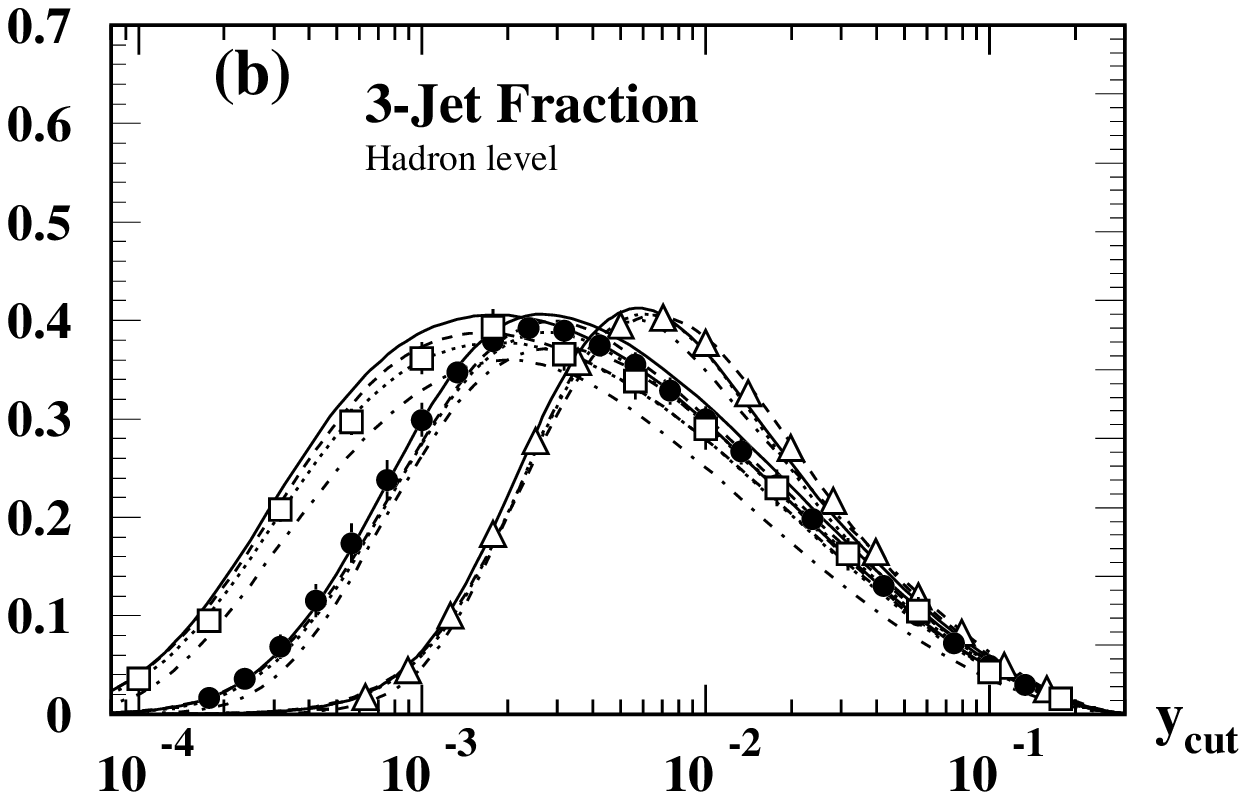,width=7cm}}
\centerline{\psfig{figure=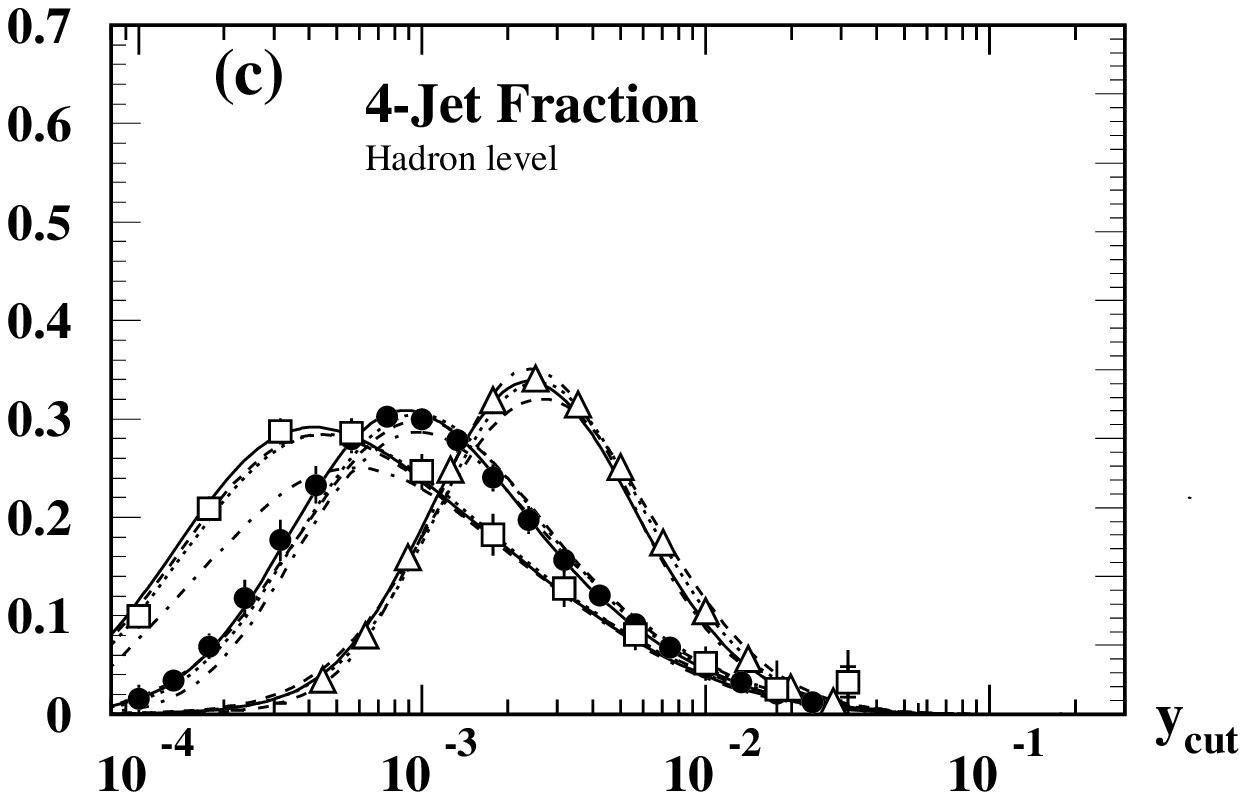,width=7cm} 
            \psfig{figure=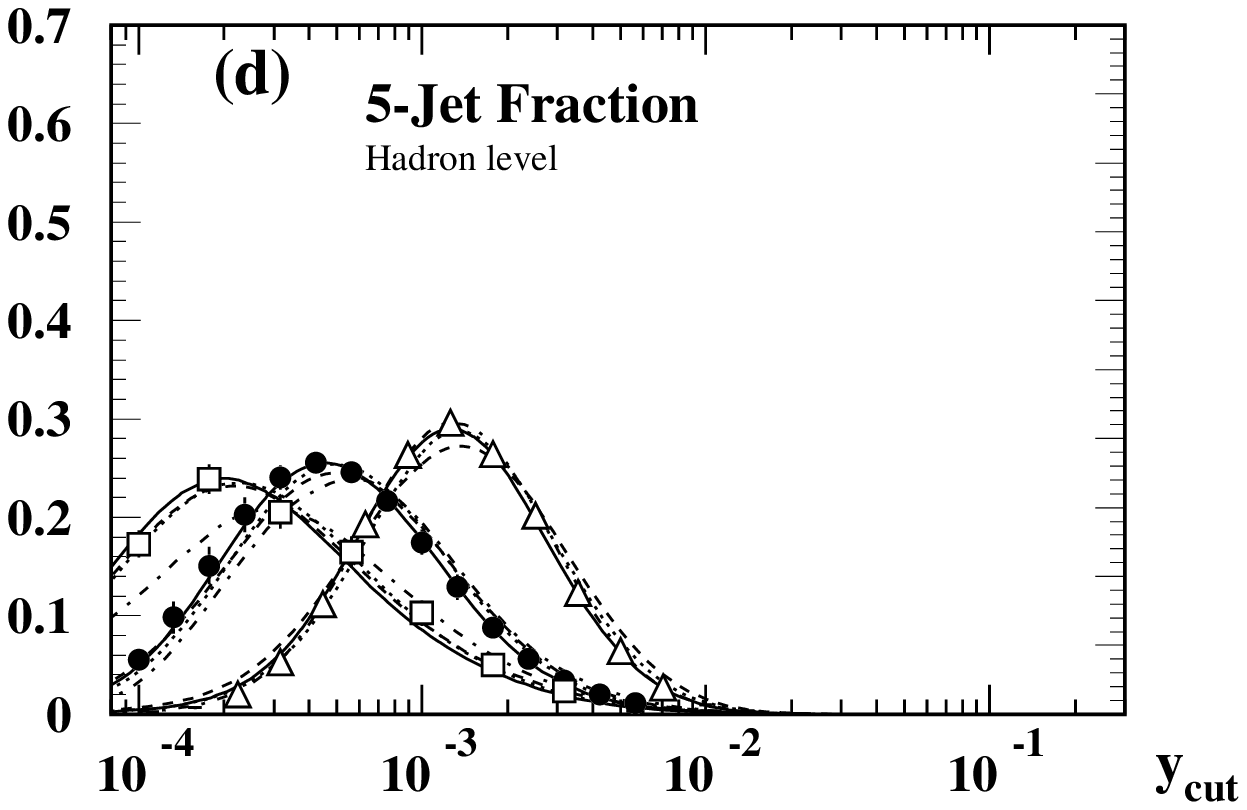,width=7cm}}
\caption{\label{fig-jetfractions}
         2-, 3-, 4-,  and 5-jet fractions for the Durham jet algorithm at
         $\sqrt{s}=35$, 91, and 189~GeV.
}
\end{figure}

\begin{figure}
\centerline{\psfig{figure=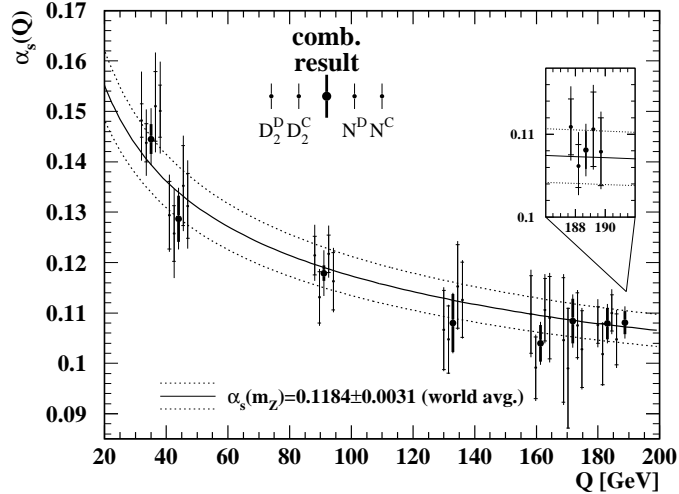,width=90mm}}
\caption{\label{fig-opaljade}
         $\alpha_S$ obtained from the differential 2-jet rate ($D_2$) and 
         the jet multiplicities ($N$) for the Durham ($^D$) and Cambridge 
         ($^C$) jet algorithms.
}
\end{figure}

Recently jet observables for the Durham and Cambridge jet algorithms have been
investigated using data of the OPAL and the JADE experiments~\cite{bib-Peterp}. 
The analysis treated the data of both experiments and estimated the errors of 
$\alpha_S$ in a similar way. Fig.~\ref{fig-jetfractions} shows the 2-, 3-, 4-,
and 5-jet fractions for the Durham jet algorithm at three different $\sqrt{s}$. 
Predictions of several models are overlaid.
The coupling strength has been obtained from fits of the matched resummed
NLLA plus second order matrix element to the differential 2-jet rate ($D_2$) 
and the jet multiplicity ($N$) of both jet finders. Fig.~\ref{fig-opaljade} 
shows the results with the world average overlaid.
Combining all eight determinations and taking correlations into account
yields $\alpha_S(m_{\mathrm{Z}}) = 0.1187 \pm 0.0010_{\mathrm{(expt)}} \ 
                                          ^{+0.0032}_{-0.0016~{\mathrm{(theo)}}}$
which is in excellent agreement with the world average~\cite{bib-asworldavg}
and has a very small total error.

\section{Flavour independence}
Finite quark masses affect the result of $\alpha_S$ determinations. In particular
bottom quark events at the Z mass yield a 7\% lower value of $\alpha_S$ if the
quark mass effect is neglected~\cite{bib-asflav}. To account for the mass effect 
for an inclusive determination which neglected this effect, the value of $\alpha_S$ 
has to be increased by about 1\%, which is covered by the typical total error.

Being precisely confirmed for the heavy charm and bottom quarks, the flavour 
independence has been scarcely tested for the light quarks at high energies. 
At $\sqrt{s}\approx 0$ evidence for the flavour independence comes from e.g.\ 
isospin invariance and approximate $SU(3)_{\mathrm{flav}}$ symmetry. The 
challenge for an investigation of the flavour independence at $\sqrt{s}=
m_{\mathrm{Z}}$ is to separate u, d, and s quark events. Using the leading 
particle effect~\cite{bib-leading} for K$^\pm$, K$^0_S$, and all kinds of
charged particles OPAL~\cite{bib-flavdep} selected events which are enriched 
differently in u, d, and s quarks and thus allows for a statistical decomposition
of the contribution of each of the three light quark flavours. $\alpha_S$ is 
determined from the charged multiplicities, $\langle N\rangle$, obtained for
each quark flavour from the decomposition, using $\langle N\rangle \sim 
\alpha_S^B \cdot \exp(C/\sqrt{\alpha_S})$ where $B$ and $C$ are known from 
QCD calculations~\cite{bib-multipli}. This yielded the preliminary ratios: 
        $\alpha_S^{\mathrm{u}}/ \alpha_S^{\mathrm{d}} = 0.88 \pm 0.08$,
        $\alpha_S^{\mathrm{s}}/ \alpha_S^{\mathrm{d}} = 0.96 \pm 0.06$, and
        $\alpha_S^{\mathrm{s}}/ \alpha_S^{\mathrm{u}} = 1.09 \pm 0.06$
which are consistent with flavour independence at a level of better than 10\%
and constitute an improvement over the previous OPAL result~\cite{bib-oldflavdep}.

\section{Summary}
QCD is in a very good shape! The fundamental properties of the theory, 
i.e.\ the running of $\alpha_S$, its flavour independence and also the 
running of the renormalised quark masses are observed and confirmed in 
experimental investigations. At the highest energies of LEP the value
of the coupling is determined to be $\alpha_S(198 {\mathrm{~GeV}})=0.109 
\pm 0.005$ (prelim.) which agrees with the expected running. The value 
of the strong coupling constant is now very precisely determined from a 
combined analysis of OPAL and JADE data covering centre-of-mass energies
from $\sqrt{s}=22$ through $189$ GeV to be $\alpha_S(m_{\mathrm{Z}}) = 
0.1187^{+0.0034}_{-0.0019}$. The flavour independence of the coupling at 
high energies is now confirmed for the light quarks at the level of better 
than 10\%.

\end{document}